%% file: main.tex
\documentclass[runningheads]{llncs} 

\usepackage{graphicx}
\usepackage[usenames,dvipsnames,svgnames,table]{xcolor}
\usepackage{mathtools}
\usepackage{amssymb,amsmath}
\usepackage{ifthen}
\usepackage{color}
\usepackage{xspace}
\usepackage{listings}
\usepackage{float}
\usepackage{makecell}
\usepackage{wrapfig}
\usepackage{lipsum}
\usepackage{array}
\usepackage[utf8]{inputenc}
\usepackage{textcomp}
\usepackage{enumitem}
\usepackage{booktabs}
\usepackage[final]{pdfpages}
\usepackage[colorlinks=true,linkcolor=blue,urlcolor=blue,citecolor=blue]{hyperref} %colorlinks=true,linkcolor=blue,urlcolor=blue,citecolor=blue
\usepackage[section]{placeins}
\usepackage{tikz}
\usetikzlibrary{decorations.pathmorphing, patterns}
\usepackage[scriptsize]{subfigure}
\usepackage[small,bf]{caption}
\usepackage{pifont}
\newcommand*\blackcircled[1]{\tikz[baseline=(char.base)]{
            \node[shape=circle,fill,inner sep=0.5pt] (char) {\textcolor{white}{#1}};}}

\usepackage{wrapfig}
\usepackage{caption}

\usepackage{pslatex}

\usepackage{tabularx}
\newcolumntype{L}[1]{>{\raggedright\arraybackslash}p{#1}}
\newcolumntype{C}[1]{>{\centering\arraybackslash}p{#1}}
\newcolumntype{R}[1]{>{\raggedleft\arraybackslash}p{#1}}

\usepackage{hyphenat}
\newboolean{showcomments}
\setboolean{showcomments}{true}
\ifthenelse{\boolean{showcomments}}
{ \newcommand{\mynote}[3]{
    \fbox{\bfseries\sffamily\scriptsize#1}
    {\small$\blacktriangleright$\textsf{\emph{\color{#3}{#2}}}$\blacktriangleleft$}}}
{ \newcommand{\mynote}[3]{}}
\newcommand{\tz}{\textsc{TrustZone}\xspace}
\newcommand{\arm}{\textsc{Arm}\xspace}
\newcommand{\optee}{\textsc{Op-Tee}\xspace}
\newcommand{\sgxspark}{\textsc{SGX-Spark}\xspace}
\newcommand{\sparksgx}{\sgxspark} % i always confuse, make one alias of the other

\begin{document}
\mainmatter

\title{Using Trusted Execution~Environments\\for Secure Stream Processing of Medical Data\\\normalsize(Case Study Paper)}
\titlerunning{Using TEEs for Secure Stream Processing of~Medical~Data}

\author{Carlos Segarra\inst{1}\orcidID{0000-0003-3455-7563}, Ricard Delgado-Gonzalo\inst{1}\orcidID{0000-0002-7183-6257}, Mathieu Lemay\inst{1}, Pierre-Louis Aublin\inst{2}\\Peter Pietzuch\inst{2}, \and Valerio Schiavoni\inst{3}\orcidID{0000-0003-1493-6603}}
\authorrunning{Segarra \emph{et al.}}
\institute{
CSEM, Neuch\^atel, Switzerland,
\email{{{cse,rdg,mly}@csem.ch}}
\and
Imperial College London, United Kingdom,
\email{{p.aublin,prp}@imperial.ac.uk}
\and
University of Neuch\^atel, Switzerland,
\email{valerio.schiavoni@unine.ch}
}
 
\maketitle
\begin{abstract}
Processing sensitive data, such as those produced by body sensors, on third-party untrusted clouds is particularly challenging without compromising the privacy of the users generating it. Typically, these sensors generate large quantities of continuous data in a streaming fashion. Such vast amount of data must be processed efficiently and securely, even under strong adversarial models. The recent introduction in the mass-market of consumer-grade processors with Trusted Execution Environments (TEEs), such as Intel SGX, paves the way to implement solutions that overcome less flexible approaches, such as those atop homomorphic encryption. We present a secure streaming processing system built on top of Intel SGX to showcase the viability of this approach with a system specifically fitted for medical data. We design and fully implement a prototype system that we evaluate with several realistic datasets. Our experimental results show that the proposed system achieves modest overhead compared to vanilla Spark while offering additional protection guarantees under powerful attackers and threat models.\footnote{This is a post-peer-reviewed, pre-copyedit version of an article published in the book \textit{"Distributed Applications and Interoperable Systems"}. The final authenticated version is available online at: \url{https://doi.org/10.1007/978-3-030-22496-7_6}}
\end{abstract}

\input{sections/introduction}
\input{sections/background}
\input{sections/architecture}

\input{sections/implementation}
\input{sections/evaluation}
\input{sections/related_work}
\input{sections/future_work}
\input{sections/conclusion}
\input{sections/acks}
\pagebreak
{\footnotesize
	\bibliographystyle{splncs04}
    \bibliography{biblio-compressed}
}

\end{document}

%% file: sections/introduction.tex
% !TEX root = ../main.tex
\section{Introduction} \label{sec:intro}
Internet of Things (IoT) devices are more and more pervasive in our lifes~\cite{Gartner2017}. The number of devices owned per user is anticipated to increase by 26$\times$ by 2020~\cite{Barbosa2017}. These devices continuously generate all large variety of continuous data. Notable examples include location-based sensors (\emph{e.g.}, GPS), inertial units (\emph{e.g.}, accelerometers, gyroscopes), weather stations, and, the focus of this paper, human-health data (\emph{e.g.}, blood pressure, heart rate, stress).

These devices usually have very restricted computing power and are typically very limited in terms of storage capacity. Hence, this continuous processing of data must be off-loaded elsewhere, in particular for storage and processing purposes. In doing so, one needs to take into account potential privacy and security threats that stem inherently from the nature of the data being generated and processed.

Cloud environments represent the ideal environment to offload such processing. They allow deployers to hand-off the maintenance of the required infrastructure, with immediate benefit for instance in terms of scale-out with the workload. 

Processing privacy-sensitive data on untrusted cloud platforms present a number of challenges. A malicious (compromised) Cloud operator could observe and leak data, if no countermeasures are taken beforehand. While there are software solutions that allow to operate on encrypted data (\emph{e.g.}, partial~\cite{Paillier1999} or full-homomorphic~\cite{Gentry2012} encryption), their current computational overhead makes  impractical in real-life scenarios~\cite{gottel2018security}.

The recent introduction into the mass market of processors with embedded trusted execution environments (TEEs), \emph{e.g.}, Intel Software Guard Extensions (SGX)~\cite{costan2016intel} (starting from processors with codename Skylake) or ARM TrustZone~\cite{trustzone}, offer a viable alternative to pure-software solutions. TEEs protect code and data against several types of attacks, including a malicious underlying OS, software bugs or threats from co- hosted applications. The application's security boundary becomes the CPU itself. The code is executed at near-native execution speeds inside enclaves of limited memory capacity. All the major Infrastructure-as-a-Service providers (Google~\cite{gceskylake}, Amazon~\cite{amazonskylake}, IBM~\cite{ibm-sgx}, Microsoft~\cite{azureconfidential}) are nowadays offering nodes with SGX processors.

We focus on the specific use case of processing data streams generated by health-monitoring wearable devices on untrusted clouds with available SGX nodes. This setting addresses the fact that algorithms for analyzing cardiovascular signals are getting more complex and computation-intensive. Thus, traditional signal-processing approaches~\cite{Kumar2016} have left the way to deep neural networks~\cite{Xiong2018,VanZaen2019}. This increase in computational expenditure has moved the processing towards centralized centers (\textit{i.e.}, the cloud) when scaling up to a large fleet of wearable devices is needed. In order to illustrate the concept, we present a system that computes in real time several metrics of the heart-rate variability (HRV) steaming from wearable sensors. While existing stream processing solutions exist~\cite{spark-whitepaper,securestreams}, they either lack support for SGX or, if they do support it, are tied to very specific programming frameworks and prevent adoption in industrial settings.

The contributions of this case-study paper are twofold. First, we design and implement a complete system that can process heart-specific signals inside SGX enclaves in untrusted clouds. Our design leverages \sgxspark, a stream processing system that exploits SGX to execute stream analytics inside TEEs (described in detail in \S\ref{sec:background}). Note that our design is flexible enough to be used with different stream processing systems (as further described later). Second, we compare the proposed system against the vanilla, non-secure Spark. Our evaluation shows that the current overhead of SGX is reasonable even for large datasets and for high-throughput workloads and that the technology is almost ready for production environments.

This paper is organized as follows. In \S\ref{sec:background}, we introduce Intel SGX, Spark, and \sgxspark. The architecture of the proposed system is presented in \S\ref{sec:architecture}, while we further provide implementation details in \S\ref{sec:implementation}. We evaluate our prototype with realistic workloads in \S\ref{sec:evaluation} for which include experimental comparisons also against the vanilla Spark. A summary of related work in the domain of (secure) stream processing is given in \S\ref{sec:related-work}. Finally, we present future work (\S\ref{sec:futurework}) before concluding in \S\ref{sec:conclusion}.

%% file: sections/background.tex
% !TEX root = ../main.tex

\section{Background}\label{sec:background}
To better understand the design and implementation details, we introduce some technical background on the underlying technologies that we leverage, as well as some of the specific features interesting for cardiac signals. In \S\ref{subsec:background:tech}, we provide background on the technical aspects exploited in the remaining of this paper, specifically describing the operating principles of Intel SGX, Spark and its secure counter-part \sgxspark. In \S\ref{subsec:background:med}, we describe the specifics of the data streams that the system has to deal with from the medical domain, such as heart-beat monitoring signals, together with the required processing that our system allows to offload on an untrusted cloud provider.

\subsection{Technical Background}\label{subsec:background:tech}
\textbf{Trusted Execution Environments and Intel SGX.}
A trusted execution environment (TEE) is an isolated area of the processor that offers code and data's confidentiality and integrity guarantees. TEEs are nowadays available in commodity CPUs, such as \arm \tz and Intel\textregistered SGX.

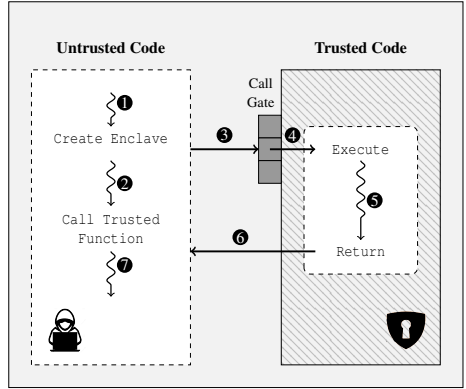
\begin{wrapfigure}{r}{.5\textwidth}
    \centering
    \input{resources/figs/sgx-principles.tex}
    \caption{\textsc{Intel SGX} execution workflow.\label{fig:sgx-principles}}
\end{wrapfigure}
In comparison with \arm \tz, SGX includes a remote attestation protocol, support multiple trusted applications on the same CPU, and its SDK is easier to program with. As mentioned earlier, all the major IaaS providers offer SGX-enabled instances on their cloud offering, hence we decided to base the design of our system on top of it. Briefly, the SGX extensions are a set of instructions and memory access extensions. These instructions enable applications to create hardware-protected areas in their address space, also known as, \emph{enclaves}~\cite{sgx-whitepaper}. At initialization time, the content loaded is measured (via hashing) and sealed. An application using an enclave identifies itself through a remote attestation protocol and, once verified, interacts with the protected region through a call gate mechanism. In particular, Figure~\ref{fig:sgx-principles} breaks down the typical execution workflow of SGX applications. After the initial attestation protocol, code in the untrusted region creates an enclave and securely loads trusted code and data inside (Figure-\ding{202}). Whenever this untrusted code wants to make use of the enclave, it makes a call to a trusted function (Figure-\ding{203}, Figure-\ding{204}) that gets captured by the call gate mechanism and, after performing sanity and integrity checks (Figure-\ding{205}), gets executed (Figure-\ding{206}), the value returned (Figure-\ding{207}) and the untrusted code can resume execution (Figure-\ding{208}). The security perimeter is kept at the CPU package and, as a consequence, all other software including privileged software, OS, hypervisors or even other enclaves are prevented from accessing code and data located inside the enclave. Most notably, the systems' main memory is left untrusted and the traffic between CPU and DRAM over the protected address range is managed by the \textit{Memory Encryption Engine}~\cite{Gueron2016}.

\textbf{Spark and Spark Streaming.}
Spark is a cluster-computing framework to develop scalable, fault-tolerant, distributed applications. It builds on RDDs, resilient distributed datasets~\cite{spark-whitepaper}, a read-only collection distributed over a cluster that can be rebuilt if one partition is lost. It is implemented in \textsc{Scala} and provides bindings for \textsc{Python}, \textsc{Java}, \textsc{SQL} and \textsc{R}. \textsc{Spark Streaming}~\cite{spark-streaming} is an extension of Spark's core API that enables scalable, high-throughput, fault tolerant stream (mini-batch) processing of data streams~\cite{spark-streaming-documentation}. The proposed system leverages Spark Streaming to perform file-based streaming, by monitoring a filesystem interface outside the enclave.
\begin{wrapfigure}{r}{.5\textwidth}
    \centering
    \input{resources/figs/sgx-spark.tex}
    \caption{\textsc{SGX-Spark} attacker model and collaborative structure scheme.\label{fig:sgx-spark-scheme}}
\end{wrapfigure}
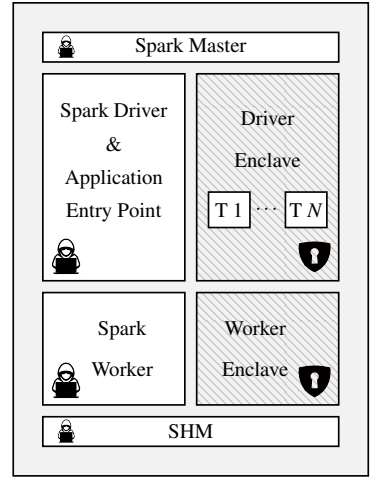
\textbf{SGX-LKL and SGX-Spark.}
\textsc{SGX-LKL}~\cite{sgx-lkl} is a library OS to run unmodified Linux binaries inside enclaves. 
Namely, it provides system support for managed runtimes, \emph{e.g.}, a full JVM. 
This feature enables the deployment of Spark, and Spark Streaming applications to leverage critical computing inside Intel SGX with minimal to no modifications to the application's code. 
\sparksgx~\cite{sgx-spark} builds on \textsc{SGX-LKL}.
It partitions the code of Spark applications to execute the sensitive parts inside SGX enclaves. 
Figure~\ref{fig:sgx-spark-scheme} depicts its architecture.
Basically, it deploys two collaborative Java Virtual Machines (JVM), one outside (Figure~\ref{fig:sgx-spark-scheme}, \textit{Spark Driver}) and one inside the enclave (Figure~\ref{fig:sgx-spark-scheme}, \textit{Driver Enclave}) for the driver, and two more for each worker deployed. 
Spark code outside the enclave accesses only encrypted data.
The communication between the JVMs is kept encrypted and is performed through the host OS shared memory. \sgxspark provides a compilation toolchain, and it currently supports the vast majority of the native Spark operators, allowing to transparently deploy and run existing Spark applications into the SGX enclaves.
%The user must only compile the source code together with \textsc{SGX-Spark}'s and, as long as the operators used are supported by the framework, execution is seamlessly deployed inside the enclave.
%\vspace{-10pt}
\subsection{Heart Rate Variability Analysis}\label{subsec:background:med}
%\vspace{-5pt}
%\vs{also stress that while we apply our architecture to this kind of streams, the architecture could be easily adapted to different use-cases.}
\begin{figure}[t]
    \centering
    \input{resources/figs/ecg-hrv.tex}
    \caption{Schematic representation of an ECG signal. It shows three normal beats and the information transferred from the sensor to the gateway. The most relevant part of the ECG wave are the R peaks and the time elapsed between them. The RR intervals together with the R peaks' timestamp are sent from the sensor to the gateway.\label{fig:ecg-hrv}}
	%\vspace{-1pt}
\end{figure}
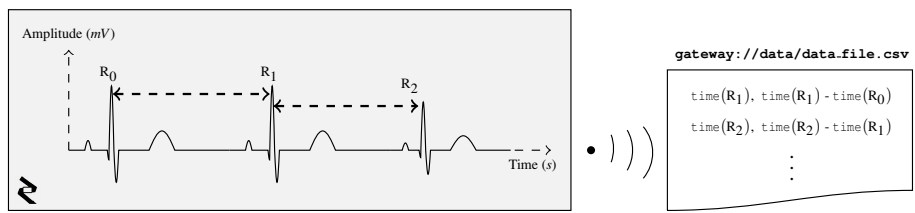
The data streams used for the evaluation and the algorithms compiled with \textsc{SGX-Spark} belong to the medical domain and motivate the real need for confidentiality and integrity. 
As further explained in \S\ref{sec:architecture}, our use case contemplates a scenario where multiple sensors track the cardiac activity of different users. The two most standard procedures for monitoring heart activity are electrocardiograms (ECG) and photoplethysmograms (PPG). An ECG measures the heart's electrical activity and is the method used by, for instance, chest bands. A PPG is an optical measure of the evolution of blood volume over time and is the method used by wrist-based sensors~\cite{Parak2015}.
%through the recording of the heart's electrical activity, e.g a band with electrodes placed on the user's chest, or (2) through an optical approximation to detect blood volume changes.~\footnote{This procedure relies on a photoplethysmogram (PPG) and is the technology used by all the wrist-based HR monitors.~\cite{ppg}} 
Both procedures lead to an approximation of R peaks' timestamps and the intervals between them (RR intervals). 
%from which the intervals between R peaks are specially relevant. 
%These elapsed times are referred to as RR intervals and are illustrated in Figure~\ref{fig:ecg-hrv}. 
The generation of the approximated diagram and the time measures are done inside the sensor.
Figure~\ref{fig:ecg-hrv} depicts a schematic representation of an ECG and the values streamed from the sensor to the gateway: R peak's timestamps and RR intervals. 
%As a consequence, the information streamed to the system is a time stamp together with the last RR interval value.
With healthy individuals' heart rate (HR) averaging between 60 to 180 beats per minute (bpm), the average throughput per client is between 23 and 69 bytes per second.
%\vs{can you mention what is the average throughput (data per second) that these typical sensors generate? - DONE}
An interesting use case of RR processing, besides HR approximation, is the study of Heart Rate Variability (HRV). 
HRV~\cite{hrv} is the variation in the time intervals between heartbeats and it has been proven to be a predictor of myocardial infarction.
Finally, despite the proposed system being specifically designed for streams with these data features, its modular design (as we later describe in \S\ref{sec:architecture}) makes it easy to adapt to other use-cases.

%% file: resources/figs/sgx-principles.tex
\resizebox{\linewidth}{!}{
\begin{tikzpicture}

    % Main outline
    %\draw[fill=green1] (0,0) rectangle (10, 1) node[pos=.5] {Privileged System Code, OS, VMM};
    \draw[fill=gray!10] (0, 0) rectangle (10, 8.5);

    % Unsecure Part
    \node[align=center] at (2.25, 7.5) {\textbf{Untrusted Code}};
    \draw[dashed, fill=white] (0.5, 0.5) rectangle (4, 7.0);
    \node at (1.25, 1.25) {\includegraphics[width=25pt]{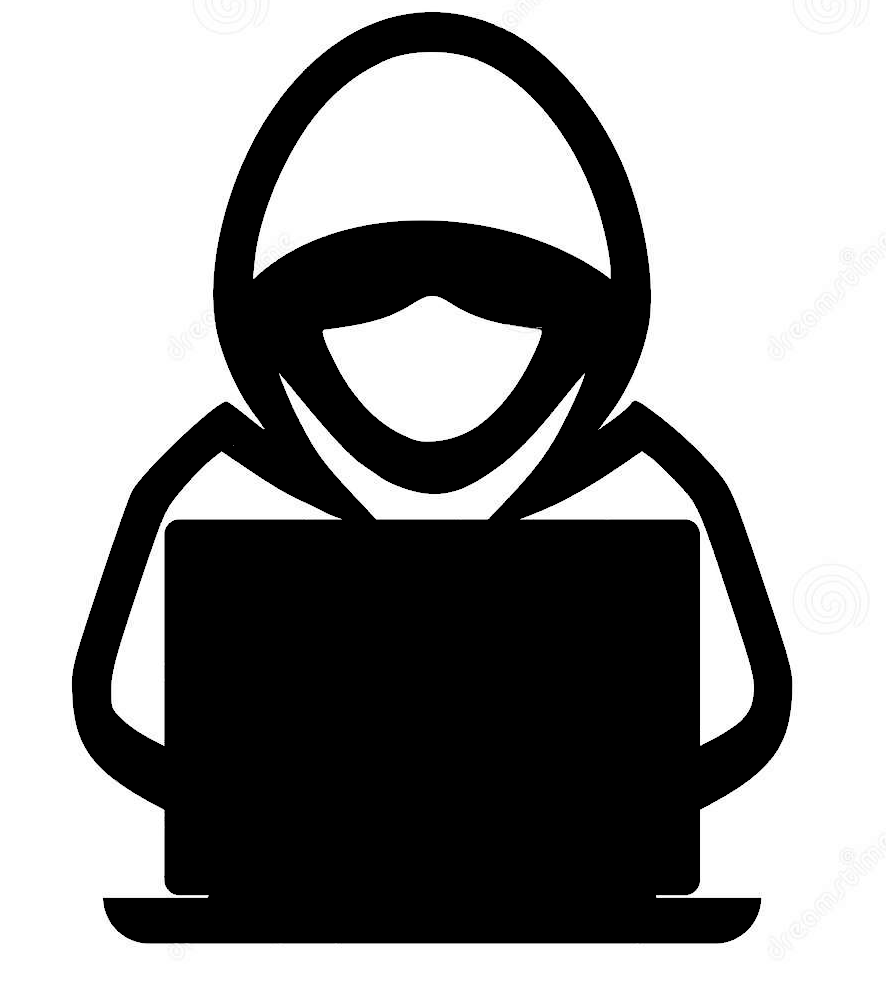}};
    \draw[->,thick, decorate,decoration={snake, post=lineto, post length=1mm}] (2.25, 6.5) -- (2.25, 5.75) node[pos=.3,anchor=west] {\blackcircled{1}};
    \node[align=center] at (2.25, 5.5) {\texttt{Create Enclave}};
    \draw[->,thick, decorate,decoration={snake, post=lineto, post length=1mm}] (2.25, 5) -- (2.25, 4) node[pos=.5,anchor=west] {\blackcircled{2}};
    \node[align=center] at (2.25, 3.5) {\texttt{Call Trusted} \\ \texttt{Function}};
    \draw[->,thick, decorate,decoration={snake, post=lineto, post length=1mm}] (2.25, 3) -- (2.25, 2) node[pos=.3,anchor=west] {\blackcircled{7}};
    \draw[->, very thick] (4, 5.25) -- (5.5, 5.25) node[pos=.5,anchor=south] {\blackcircled{3}};

    % Secure Part
    \node[align=center] at (7.75, 7.5) {\textbf{Trusted Code}};
    \draw[fill=white,pattern=north west lines,pattern color=gray!50] (6, 0.5) rectangle (9.5, 7.0);
    \node at (8.75, 1.25) {\includegraphics[width=25pt]{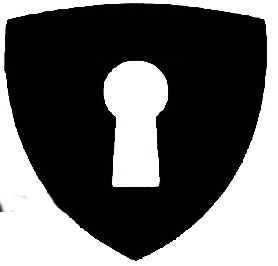}};
    \node[align=center] at (5.55, 6.5) {\small{Call} \\ \small{Gate}};
    \draw[fill=gray!80] (5.5, 5.5) rectangle (6, 6.0);
    \draw[fill=gray!80] (5.5, 5.0) rectangle (6, 5.5);
    \draw[fill=gray!80] (5.5, 4.5) rectangle (6, 5.0);
    \draw[rounded corners, dashed, fill=white] (6.5, 2.5) rectangle (9, 5.75);
    \draw[->, very thick] (5.75, 5.25) -- (6.75, 5.25) node[pos=.5,anchor=south] {\blackcircled{4}};
    \node[align=center] at (7.75, 5.25) {\texttt{Execute}};
    \draw[->,thick, decorate,decoration={snake, post=lineto, post length=4mm}] (7.75, 5) -- (7.75, 3.25) node[pos=.5,anchor=west] {\blackcircled{5}};
    \node[align=center] at (7.75, 3) {\texttt{Return}};
    \draw[->, very thick] (6.75, 3) -- (4, 3) node[pos=.6, anchor=south] {\blackcircled{6}};
\end{tikzpicture}}

%% file: resources/figs/sgx-spark.tex
\resizebox{.4\textwidth}{!}{
\begin{tikzpicture}
    % Colors definition: latexcolor.com
    \definecolor{ashgrey}{rgb}{0.7, 0.75, 0.71}
    \definecolor{x11gray}{rgb}{0.75, 0.75, 0.75}

    % Main outline
    \draw[fill=gray!10] (0,0) rectangle (3, 4);

    % First Layer
    % Shared Memory
    \draw[fill=white] (0.25, 0.25) rectangle (2.75, 0.5) node[pos=.5] {\tiny{SHM}};
    \node at (0.45, 0.375) {\includegraphics[width=5pt]{resources/img/hacker.png}};

%    \draw[fill=white] (0.25, 0.25) rectangle (1.45, 0.825) node[pos=.5] {\tiny{OS}};;
%    \node at (0.45, 0.45) {\includegraphics[width=8pt]{resources/img/hacker.png}};
%    \draw[fill=white, pattern=north west lines,pattern color=gray!50] (1.55, 0.25) rectangle (2.75, 0.825) node[pos=.5] {\tiny{SGX}};
%    \node at (2.55, 0.45) {\includegraphics[width=8pt]{resources/img/intel-sgx.png}};

    % Second Layer
    \draw[fill=white] (0.25, 0.6) rectangle (1.45, 1.55) node[pos=.5, xshift=2pt, align=center] {\tiny{Spark} \\[-2pt] \tiny{Worker}};
    \node at (0.45, 0.8) {\includegraphics[width=8pt]{resources/img/hacker.png}};
    \draw[fill=white, pattern=north west lines,pattern color=gray!50] (1.55, 0.6) rectangle (2.75, 1.55) node[pos=.5, xshift=-3pt, align=center] {\tiny{Worker} \\[-2pt] \tiny{Enclave}};
    \node at (2.55, 0.8) {\includegraphics[width=8pt]{resources/img/intel-sgx.png}};

    % Third Layer
    \draw[fill=white] (0.25, 1.65) rectangle (1.45, 3.4);
    \node[align=center] at (0.85, 2.65) {\tiny{Spark Driver} \\[-4pt] \tiny{\&} \\[-4pt] \tiny{Application} \\[-4pt] \tiny{Entry Point}};
    \node at (0.45, 1.85) {\includegraphics[width=8pt]{resources/img/hacker.png}};
    \draw[fill=white, pattern=north west lines,pattern color=gray!50] (1.55, 1.65) rectangle (2.75, 3.4);
    \node[align=center] at (2.15, 2.85) {\tiny{Driver} \\[-2pt] \tiny{Enclave}};
    \node at (2.55, 1.85) {\includegraphics[width=8pt]{resources/img/intel-sgx.png}};
    % Spark Tasks
    \draw[fill=white] (1.65, 2.1) rectangle (2.0, 2.4) node[pos=.5] {\tiny{T $1$}}; 
    \node at (2.15, 2.25) {\tiny{$\cdots$}};
    \draw[fill=white] (2.3, 2.1) rectangle (2.65, 2.4) node[pos=.5] {\tiny{T $N$}}; 

    % Spark Master
    \draw[fill=white] (0.25, 3.5) rectangle (2.75, 3.75) node[pos=.5] {\tiny{Spark Master}};
    \node at (0.45, 3.625) {\includegraphics[width=5pt]{resources/img/hacker.png}};

\end{tikzpicture}}

%% file: resources/figs/ecg-hrv.tex
\resizebox{\linewidth}{!}{
\begin{tikzpicture}
    % Colors definition
    \pgfdeclaredecoration{single pulse}{initial}{
    \state{initial}[width=\pgfdecoratedinputsegmentlength]
    {%
        % Initial Line
        \pgfpathlineto{\pgfpoint{0.1*\pgfdecoratedinputsegmentlength}{0mm}}%    
        % P Peak
        \pgfpathsine{\pgfpoint{0.2\pgfdecorationsegmentlength}{0.15\pgfdecorationsegmentamplitude}}%
        \pgfpathcosine{\pgfpoint{0.2\pgfdecorationsegmentlength}{-0.15\pgfdecorationsegmentamplitude}}%
        % P - Q Line
        \pgfpathlineto{\pgfpoint{0.6\pgfdecorationsegmentamplitude}{0mm}}%
        % Q Valley
        \pgfpathsine{\pgfpoint{0.1\pgfdecorationsegmentlength}{-0.15\pgfdecorationsegmentamplitude}}
        \pgfpathcosine{\pgfpoint{0.01\pgfdecorationsegmentlength}{0.15\pgfdecorationsegmentamplitude}}%
        % R Peak
        \pgfpathsine{\pgfpoint{0.15\pgfdecorationsegmentlength}{\pgfdecorationsegmentamplitude}}%
        \pgfpathcosine{\pgfpoint{0.15\pgfdecorationsegmentlength}{-\pgfdecorationsegmentamplitude}}%
        % S Valley
        \pgfpathsine{\pgfpoint{0.15\pgfdecorationsegmentlength}{-0.5\pgfdecorationsegmentamplitude}}
        \pgfpathcosine{\pgfpoint{0.15\pgfdecorationsegmentlength}{0.5\pgfdecorationsegmentamplitude}}%
        % S to T line
        \pgfpathlineto{\pgfpoint{1.25\pgfdecorationsegmentamplitude}{0mm}}%
        % T Peak
        \pgfpathsine{\pgfpoint{0.8\pgfdecorationsegmentlength}{0.3\pgfdecorationsegmentamplitude}}%
        \pgfpathcosine{\pgfpoint{0.8\pgfdecorationsegmentlength}{-0.3\pgfdecorationsegmentamplitude}}%
        % Last Line
        \pgfpathlineto{\pgfpointdecoratedinputsegmentlast}%
    }
    \state{final}{}%
    }

    \fill[gray!10, draw=black] (-0.75, -0.75) rectangle (6.25, 1.75);
    \node at (-0.5, -0.5) {\includegraphics[width=10pt]{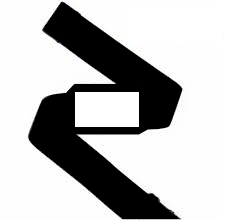}};
    %\node at (5.5, 1.5) {\textbf{\small{Sensor}}};

    \draw[->, dashed] (5.55,0) -- (6,0) node[pos=.5, anchor=north] {\tiny{Time ($s$)}};
    \draw[->, dashed] (0,0) -- (0,1.25) node[anchor=south] {\tiny{Amplitude ($mV$)}};
    \draw[decoration={single pulse,amplitude=8mm,segment length=2mm},decorate] (0,0) -- (2,0);
    \draw[decoration={single pulse,amplitude=8mm,segment length=2mm},decorate] (2,0) -- (4,0);
    \draw[decoration={single pulse,amplitude=6mm,segment length=2mm},decorate] (4,0) -- (5.5,0);
    \draw[<->, dashed, thick] (0.55, 0.7) -- (2.5, 0.7);
    \draw[<->, dashed, thick] (2.55, 0.55) -- (4.35, 0.55);
    \node[align=center] at (0.5, 0.95) {\tiny{$\text{R}_0$}};
%    \node[align=center] at (2.2, 0.3) {\tiny{$\text{P}_1$}};
%    \node[align=center] at (2.45, -0.2) {\tiny{$\text{Q}_1$}};
%    \node[align=center] at (2.7, -0.55) {\tiny{$\text{S}_1$}};
%    \node[align=center] at (3.15, 0.38) {\tiny{$\text{T}_1$}};
    \node[align=center] at (2.5, 0.95) {\tiny{$\text{R}_1$}};
    \node[align=center] at (4.25, 0.8) {\tiny{$\text{R}_2$}};

    % Signal
    \fill[black] (6.5, 0) circle (0.05);
    \draw (6.75, -0.15) arc (-20:20:0.5);
    \draw (6.95, -0.25) arc (-30:30:0.5);
    \draw (7.15, -0.35) arc (-40:40:0.5);

    % Data file
    \node[align=center] at (9, 1.2) {\tiny{\textbf{\texttt{gateway://data/data\_file.csv}}}};
    \draw (7.45, -0.75) -- (7.45, 1) -- (10.5,1) -- (10.5, -0.5) to[out=180, in=0] (7.45, -0.75);
    \node[align=center] at (9, 0.2) {\tiny{\texttt{time}$\left(\text{R}_1\right)$, \xspace\texttt{time}$\left(\text{R}_1\right)$ - \texttt{time}$\left(\text{R}_0\right)$} \\ \tiny{\texttt{time}$\left(\text{R}_2\right)$, \xspace\texttt{time}$\left(\text{R}_2\right)$ - \texttt{time}$\left(\text{R}_1\right)$} \\ \tiny{\textbf{$\vdots$}}};
\end{tikzpicture}}

%% file: sections/architecture.tex
% !TEX root = ../main.tex

\section{Architecture} \label{sec:architecture}
\begin{figure}[t]
    \centering
    \input{resources/figs/system-architecture.tex}
    \caption{(Left) Schematic of the system's main architecture. A set of clients bidirectionally stream data to a remote server. The interaction is done via a filesystem interface. On the server side, \sgxspark performs secure processing using different HRV analysis algorithms. (Right) Breakdown of a packaged client: it includes a \texttt{sensor} and gateway that wrap four different microservices (\textsc{mqtt} broker, \texttt{mqtt-subscriber}, \texttt{consumer}, \texttt{producer}) to interact with the remote end. \label{fig:system-architecture}}
\end{figure}
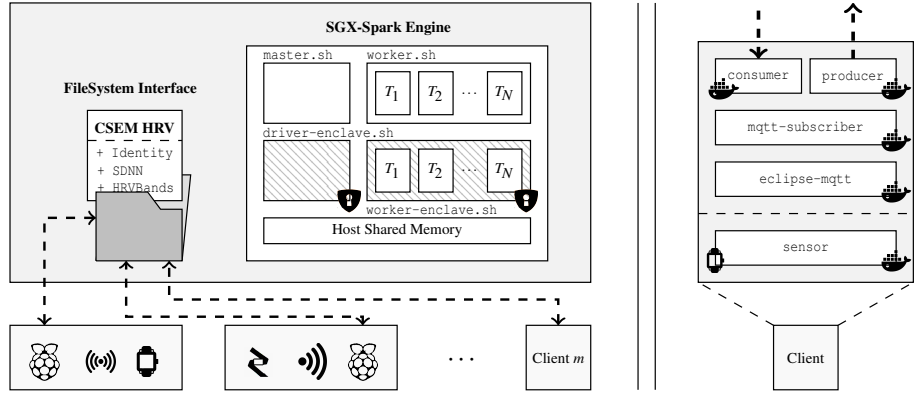

The architecture of the proposed system is depicted in Figure~\ref{fig:system-architecture}. It is composed of a server-side component which executes on untrusted machines (\emph{e.g.}, nodes on the cloud), where Intel SGX is available. The  clients are distributed among remote locations.  Each client is a sensor generating samples, and a gateway aggregating and sending them periodically every \emph{n} seconds to the cloud-based component. Similarly, clients fetch the results at fixed time intervals (\emph{i.e.}, every 5 seconds in our deployments). The interaction between the clients and the server-side components of the system happens over a filesystem interface. Each client data stream is processed in parallel by the \sgxspark job. In the reminder, we further detail these components.

\subsection{Server-side}
The server-side component is made by three different modules: a filesystem interface, the \sgxspark engine, and a set of algorithms to analyze HRV. The filesystem interface acts as a landing point for the batches of data generated by each client. It is monitored by the \sgxspark engine. Currently, it is mounted and unmounted, respectively at start-up time and upon the shutdown of the service. The streaming engine and the pool of algorithms are compiled together by the same toolchain, yet they are independent. The \textsc{Spark} engine (deployed in standalone mode) executes: the master process, the driver process, and an arbitrary number of workers. In the case of \sgxspark jobs, two JVMs are deployed per driver and worker process: one inside an enclave and one outside. The communication between JVMs is kept encrypted and is done through the host OS shared memory (see Figure~\ref{fig:sgx-spark-scheme}). For each JVM pair, \sgxspark will initialize a new enclave. The specific algorithm that the system will execute is currently set at start-up time, although several concurrent ones can be executed, each yielding separated results. 

\subsection{Clients}
The client is a combination of: (1) a data generator that simulates a sensor and (2) a gateway that interacts with the remote end. The data generator streams RR intervals. These samples are gathered by the gateway, which stacks and sends them for processing in a file-based streaming fashion. The typical size of these batches is in the 230---690 Bytes range. Each gateway is composed by: a message broker that handles the samples, a service that handles data pre-processing and batch sending, and a fetcher that directly \textit{greps} from the server's filesystem.

\subsection{Threat Model}
We assume that the communication between the gateway and the filesystem is kept protected (\emph{e.g.}, encrypted) using secure transfer protocols (more in Section~\ref{sec:implementation}). Given this assumption, the threat model is the same as typical systems that rely on \textsc{SGX}. Specifically, we assume the system software is untrusted. Our security perimeter only includes the internals of the CPU package. The trusted computing base is Intel's microcode as well as and the code loaded at the enclave, which can be measured and integrity can be checked. We assume that in our case the client package is trusted and tamper-proof. We focus on protecting the areas \emph{outside} user's control. However, if the client package is deployed in, for instance, a \textsc{Raspberry Pi}, the Trusted Computing Base (TCB) could be further reduced using \arm\tz and \optee~\cite{optee}.

\subsection{Known Vulnerabilities}
As for the known vulnerabilities, \textsc{SGX} (in particular the memory encryption engine) is not designed to be an oblivious RAM. As a consequence and adversary can perform traffic analysis attacks~\cite{Gueron2016}. Moreover, side-channel attacks~\cite{sgx-sidechannel} and speculative execution attacks (\textit{Spectre}-like~\cite{sgx-spectre} and \textit{Foreshadow}~\cite{foreshadow}) have still successful against enclaves and will require in-silicon fixes.

%% file: resources/figs/system-architecture.tex
\resizebox{\linewidth}{!}{
\begin{tikzpicture}
    % Colors definition: latexcolor.com
    \definecolor{ashgrey}{rgb}{0.7, 0.75, 0.71}
    \definecolor{x11gray}{rgb}{0.75, 0.75, 0.75}
    \definecolor{metal}{rgb}{0.43, 0.5, 0.5}

    % Main outline
    % Background
    \draw[fill=gray!10] (0, 2.75) rectangle (6.75, 6.0);

%    % Client Side
%    % Client boxes line
%    \draw[fill=metal!50] (0,1) rectangle (0.75, 1.75) node[pos=.5] {\tiny{Client $1$}};
%    \draw[fill=metal!50] (1,1) rectangle (1.75, 1.75) node[pos=.5] {\tiny{Client $2$}};
%    \draw[fill=metal!50] (2,1) rectangle (2.75, 1.75) node[pos=.5] {\tiny{Client $3$}};
%    \node at (3.425, 1.35) {$\cdots$};
%    \draw[fill=metal!50] (4,1) rectangle (4.75, 1.75) node[pos=.5] {\tiny{Client $i$}};
%    \node at (5.425, 1.35) {$\cdots$};
%    \draw[fill=metal!50] (6,1) rectangle (6.75, 1.75) node[pos=.5] {\tiny{Client $m$}};
%    \draw[fill=metal!50] (8.875,1) rectangle (9.625, 1.75) node[pos=.5] {\tiny{Client}};
%    % Lines to filesystem
%    \draw[<->, dashed, thick] (0.3725, 1.75) -- (0.3725, 3.5) -- (1, 3.5);
%    \draw[<->, dashed, thick] (1.1, 1.75) -- (1.1, 3);
%    \draw[<->, dashed, thick] (2.3725, 1.75) -- (2.3725, 2) -- (1.35, 2) -- (1.35, 3);
%    \draw[<->, dashed, thick] (4.3725, 1.75) -- (4.3725, 2.2) -- (1.6, 2.2) -- (1.6, 3);
%    \draw[<->, dashed, thick] (6.3725, 1.75) -- (6.3725, 2.4) -- (1.85, 2.4) -- (1.85, 3);
%    \draw[dashed, thin] (8.875, 1.75) -- (8, 2.75);
%    \draw[dashed, thin] (9.625, 1.75) -- (10.5, 2.75);

    % Client Side
    % Client boxes line
    % Client 1
    \draw[fill=gray!5] (0,1.5) rectangle (2, 2.25);
    \node at (1.6, 1.825) {\includegraphics[width=10pt]{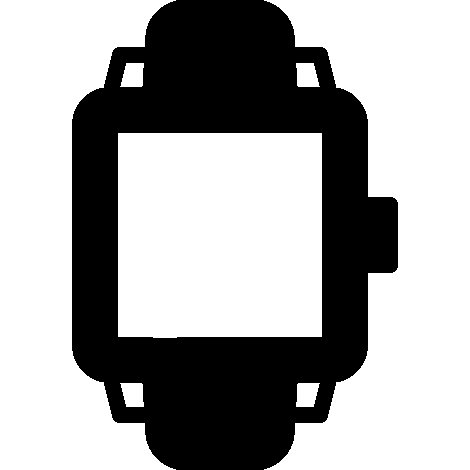}};
    \node at (1.05, 1.825) {\includegraphics[width=10pt]{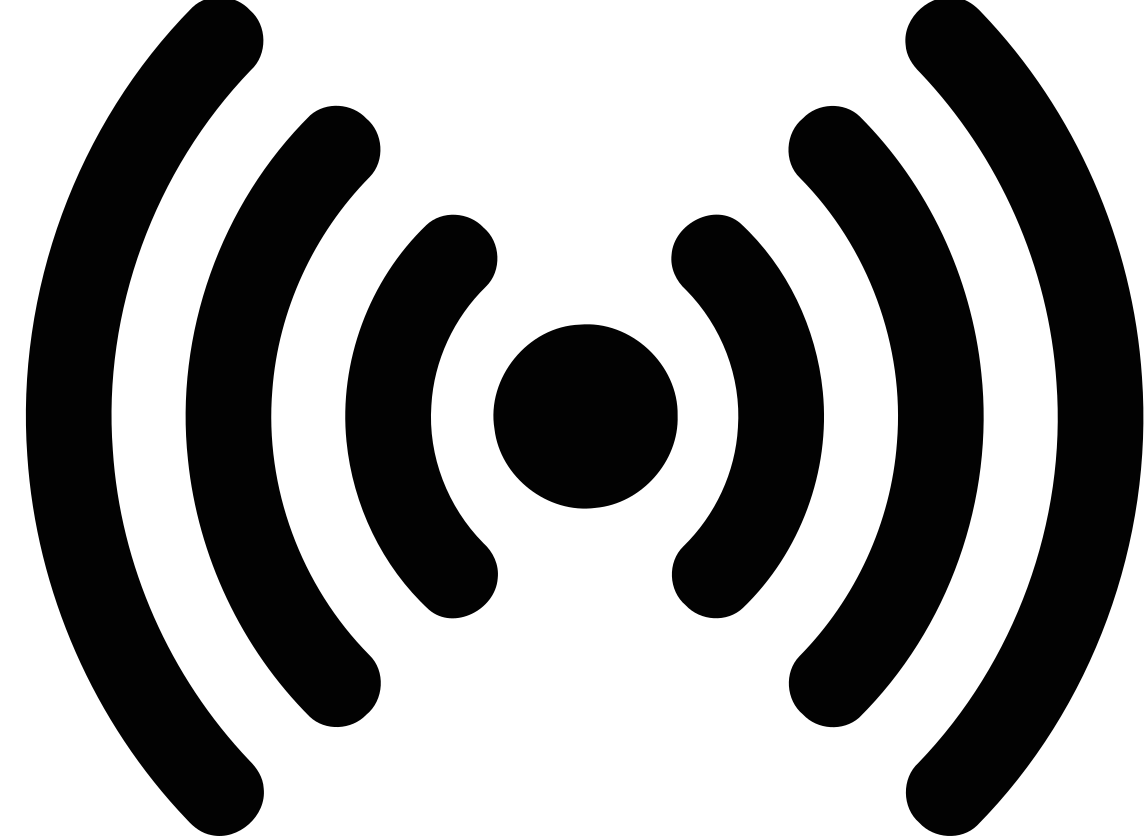}};
    \node at (0.4, 1.825) {\includegraphics[width=15pt]{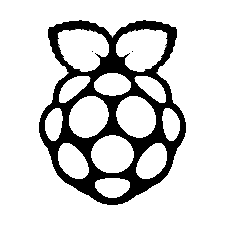}};
    % Client 2
    \draw[fill=gray!5] (2.5,1.5) rectangle (4.5, 2.25);
    \node at (2.9, 1.825) {\includegraphics[width=10pt]{resources/img/hrband.png}};
    \node at (3.5, 1.825) {\includegraphics[width=10pt]{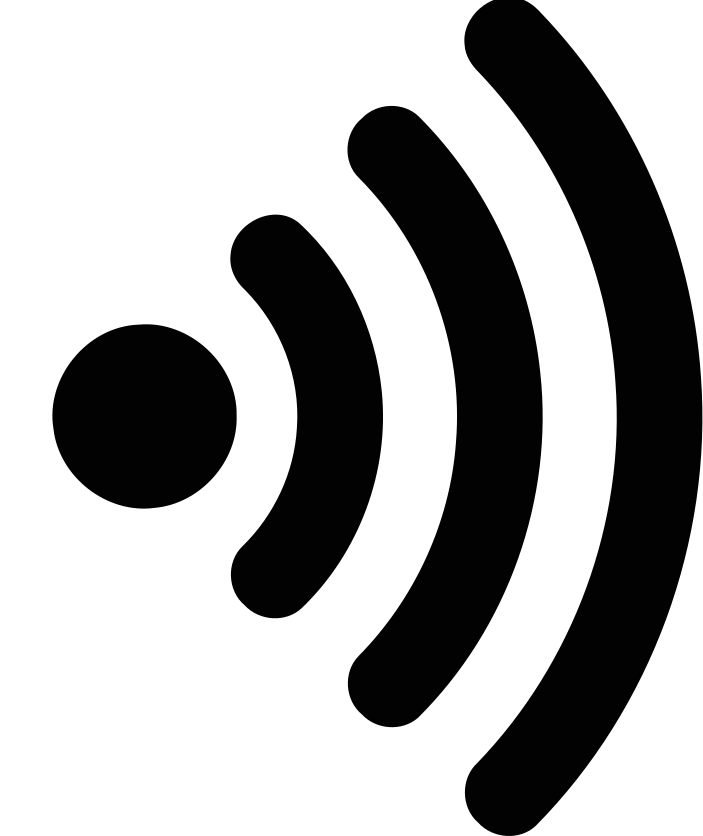}};
    \node at (4.1, 1.825) {\includegraphics[width=15pt]{resources/img/raspberry.png}};
    % Client 3
    \node at (5.265, 1.85) {$\cdots$};
    \draw[fill=gray!5] (6,1.5) rectangle (6.75, 2.25) node[pos=.5] {\tiny{Client $m$}};
    \draw[fill=gray!5] (8.875,1.5) rectangle (9.625, 2.25) node[pos=.5] {\tiny{Client}};
    % Lines to filesystem
    \draw[<->, dashed, thick] (0.4, 2.25) -- (0.4, 3.5) -- (1, 3.5);
    \draw[<->, dashed, thick] (4.1, 2.25) -- (4.1, 2.35) -- (1.35, 2.35) -- (1.35, 3);
    \draw[<->, dashed, thick] (6.3725, 2.25) -- (6.3725, 2.55) -- (1.85, 2.55) -- (1.85, 3);
    \draw[dashed, thin] (8.875, 2.25) -- (8, 2.75);
    \draw[dashed, thin] (9.625, 2.25) -- (10.5, 2.75);

    % Server Side
    % FileSystem Logo
    \draw[fill=x11gray!50] (1.0, 3) -- (1.1, 3.9) -- (1.6, 3.9) -- (1.75, 4.0) -- (2.1, 4.0) -- (2.0, 3);
    %\draw[fill=x11gray] (1.0, 3) -- (1.0, 3.8) -- (1.6, 3.8) -- (1.75, 3.6) -- (2.0, 3.6) -- (2.0, 3) -- (1.0, 3);
    \node at (1.4, 5.0) {\text{\tiny{\textbf{FileSystem Interface}}}};
    % SGX Spark
    % Main Outline
    \draw[fill=white] (2.75, 3) rectangle (6.25, 5.5);
    \node at (4.4, 5.7) {\text{\textbf{\tiny{SGX-Spark Engine}}}};
    % SHM
    \draw (2.95, 3.2) rectangle (6.05, 3.5) node[pos=.5] {\tiny{Host Shared Memory}};
    % Driver
    \draw[pattern=north west lines,pattern color=gray!50] (2.95, 3.7) rectangle (3.95, 4.4); 
    \node at (3.7, 4.5) {\tiny{\texttt{driver-enclave.sh}}};
    \node at (3.95, 3.7) {\includegraphics[width=8pt]{resources/img/intel-sgx.png}};
    \draw (2.95, 4.6) rectangle (3.95, 5.3); 
    \node at (3.35, 5.4) {\tiny{\texttt{master.sh}}};
    % Worker
    \draw[pattern=north west lines,pattern color=gray!50] (4.15, 3.7) rectangle (6.05, 4.4);
    \node at (4.9, 3.6) {\tiny{\texttt{worker-enclave.sh}}};
    \draw (4.15, 4.6) rectangle (6.05, 5.3);
    \node at (4.55, 5.4) {\tiny{\texttt{worker.sh}}};
    % Tasks Enclave
    \draw[fill=white] (4.25, 3.8) rectangle (4.65, 4.3) node[pos=.5] {\tiny{$T_1$}};
    \draw[fill=white] (4.75, 3.8) rectangle (5.15, 4.3) node[pos=.5] {\tiny{$T_2$}};
    \node at (5.35, 4.05) {\tiny{$\cdots$}};
    \draw[fill=white] (5.55, 3.8) rectangle (5.95, 4.3) node[pos=.5] {\tiny{$T_N$}};
    \node at (6, 3.7) {\includegraphics[width=8pt]{resources/img/intel-sgx.png}};
    % Tasks Outside Enclave
    \draw (4.25, 4.7) rectangle (4.65, 5.2) node[pos=.5] {\tiny{$T_1$}};
    \draw (4.75, 4.7) rectangle (5.15, 5.2) node[pos=.5] {\tiny{$T_2$}};
    \node at (5.35, 4.95) {\tiny{$\cdots$}};
    \draw (5.55, 4.7) rectangle (5.95, 5.2) node[pos=.5] {\tiny{$T_N$}};

    % CSEM's Toolbox
    \draw[fill=white] (0.9, 3.7) rectangle (2.0, 4.75);
    \node at (1.45, 4.55) {\textbf{\tiny{CSEM HRV}}};
    \draw[dashed] (0.9, 4.4) -- (2.0, 4.4);
    \node at (1.45, 4.25) {\texttt{\tiny{+ Identity}}};
    \node at (1.28, 4.05) {\texttt{\tiny{+ SDNN}}};
    \node at (1.46, 3.85) {\texttt{\tiny{+ HRVBands}}};
    \draw[fill=x11gray] (1.0, 3) -- (1.0, 3.8) -- (1.6, 3.8) -- (1.75, 3.6) -- (2.0, 3.6) -- (2.0, 3) -- (1.0, 3);

    % Client Expansion
    % Separator
    \draw (7.3, 1.5) -- (7.3, 6);
    \draw (7.5, 1.5) -- (7.5, 6);
    % Client itself
    \draw[fill=gray!10] (8, 2.75) rectangle (10.5, 5.55);
    %\draw (9.25, 1.3725) circle (0.5);
    \draw[dashed] (8, 3.55) -- (10.5, 3.55);
    \draw[fill=white] (8.2, 2.95) rectangle (10.3, 3.35) node[pos=.5] {\tiny{\texttt{sensor}}};
    \node at (10.3, 3.00) {\includegraphics[width=10pt]{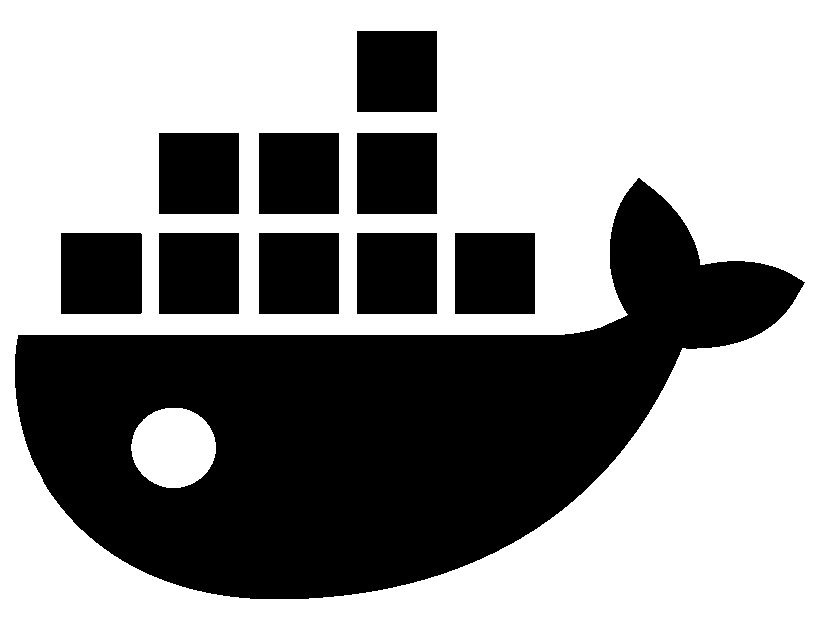}};
    \node at (8.2, 3.00) {\includegraphics[width=8pt]{resources/img/smartwatch.png}};
    \draw[fill=white] (8.2, 3.75) rectangle (10.3, 4.15) node[pos=.5] {\tiny{\texttt{eclipse-mqtt}}};
    \node at (10.3, 3.80) {\includegraphics[width=10pt]{resources/img/docker.png}};
    \draw[fill=white] (8.2, 4.35) rectangle (10.3, 4.75) node[pos=.5] {\tiny{\texttt{mqtt-subscriber}}};
    \node at (10.3, 4.40) {\includegraphics[width=10pt]{resources/img/docker.png}};
    \draw[fill=white] (8.2, 4.95) rectangle (9.2, 5.35) node[pos=.5] {\tiny{\texttt{consumer}}};
    \node at (8.3, 5.00) {\includegraphics[width=10pt]{resources/img/docker.png}};
    \draw[fill=white] (9.3, 4.95) rectangle (10.3, 5.35) node[pos=.5] {\tiny{\texttt{producer}}};
    \node at (10.3, 5.00) {\includegraphics[width=10pt]{resources/img/docker.png}};
    \draw[->, very thick, dashed] (9.8, 5.35) -- (9.8, 6);
    \draw[<-, very thick, dashed] (8.7, 5.35) -- (8.7, 6);

\end{tikzpicture}}

%% file: sections/implementation.tex
% !TEX root = ../main.tex
\section{Implementation}\label{sec:implementation}
This section presents the further implementation details. To stress-test our evaluation, we replaced real sensors with synthetic data generators. Additionaly, we deploy a large number of Docker containers~\cite{docker-container} to mimic a fleet of concurrent clients.

\subsection{Server-side}
We rely on the original \sgxspark implementation, and we only modify it to support a different in-enclave code deployment path, so that the \texttt{.jar} archive is available inside the enclaves and the shared memory. The application code is implemented in the Scala programming language~\cite{scala-language}.
Applications must adhere to the RDD API~\cite{rdd-programming-guide} to be usable inside the SGX enclaves. 
We use \sgxspark via Structured Streaming jobs, and  must also adhere to the same API. 
We have implemented two state-of-the-art HRV analysis algorithms, namely \texttt{SDNN} and \texttt{HRVBands}~\cite{hrv-metrics}.
The SDNN algorithm measures the standard deviation of NN (RR in our case) intervals. HRVBands performs frequency domain calculations: high-frequency (HF) power, low-frequency (LF) power and HF to LF ratio. For the sake of performance comparison, we also include results using an identity algorithm, simply reading the input data stream and outputting it. The implementation of these algorithms rely on basic Spark Streaming operators, and their corresponding Scala implementations. We use the file-based data stream input for \textsc{Spark} streaming.\footnote{\url{https://spark.apache.org/docs/2.2.0/api/java/org/apache/spark/streaming}} 

\subsection{Clients}
Clients correspond to body-sensors strapped to the body of a user. These are connected to a gateway,  (\emph{e.g.}, a Raspberry Pi) packaged together. Our implementation decouples the clients into into five different microservices (see Figure~\ref{fig:system-architecture}, right). For evaluation purposes, the \texttt{sensor} is a \textsc{Python} service that generates random RR intervals. These are published into the \textsc{MQTT} queue~\cite{mqtt-protocol,mqtt-eclipse} following a uniform time distribution. The gateway is composed by a \textsc{MQTT} queue and broker service. We rely on \texttt{eclipse-mosquitto}\footnote{\url{https://hub.docker.com/_/eclipse- mosquitto/}}, a \texttt{mqtt-sub} service that subscribes to the specific topic and generates data files with several samples, and a \texttt{producer} and \texttt{consumer} services that interact with the remote filesystem. These components are implemented in Python, and consist of $888$ Lines of Code (LoC). Our prototype relies on Docker to facilitate the deployment of new clients, and on \texttt{docker-compose}~\cite{docker-compose} to easily group orchestrate their deployment. The communication between the client and the server happens via SSH/SecureFTP to ensure transport layer security when transferring user's data.

\subsection{Deployment}
To ease scalability and reproducibility of both server and client, deployment is orchestrated by a single script detached from both execution environments. Specifying the remote location, the \textsc{SGX-Spark} engine, the streaming algorithm and the filesystem interface are initialized either container-based or on metal. Specifying the number of simulated users and their location, a cluster of clients is dynamically started. On execution time, a Spark streaming service located in a remote server with a master process and an arbitrary number of Spark workers (or executors) interacts with a standalone Docker Swarm composed by the cluster of clients, a name discovery service and an overlay network. This architecture scales to hundreds of clients.

%% file: sections/evaluation.tex
% !TEX root = ../main.tex
\section{Evaluation} \label{sec:evaluation}
In this section, we present the experimental evaluation. We first present the evaluation settings for both the client and the server components. Then, we describe the metrics of interest on which we focus our experiments. 
Finally, we present our results. Our experiments answer the following questions: \emph{(i)} is the design of the proposed system sound? \emph{(ii)} is our implementation efficient, \emph{(iii)} what is the overhead of SGX, and \emph{(iv)} is it scalable?

\subsection{Settings}

\textbf{Clients.}
Each client (\emph{e.g.}, a body sensor in real-life) is emulated by a standalone Docker application.
We deploy them on a quad-16core (64 hardware cores) AMD EPYC 7281 with 64 GiB of RAM running Ubuntu v$18.04$ LTS (kernel 4.15.0-42-generic). 
The client containers are built and deployed using Docker (v$18.09.0$) and \texttt{docker-compose} (v$1.23.2$). 
We use \texttt{docker-machine} (v$0.16.0$) with the \texttt{virtualbox} disk image. 
Each machine hosts 20 clients, the maximum number of services supported by its local network, and it registers itself to the Swarm via a name discovery service running on another machine.
Inter-container communication rely on the \texttt{overlay} network driver.
We pull the latest images available on Docker Hub for the Consul name discovery service (v$1.4$) and the \texttt{eclipse-mosquitto} (v$1.5$) message broker.

\textbf{Server.} 
The server components run on host machines with Intel~\textregistered\xspace Xeon~\textregistered\xspace CPU E3-1270 v6 @ $3.80$~GHz with 8 cores and 64 GiB RAM. 
We use Ubuntu $16.04$ LTS (kernel 4.19.0-41900-generic) and the official Intel~\textregistered\xspace SGX driver v$2.0$~\cite{sgx-driver}, and \textsc{SGX-LKL}~\cite{sgx-lkl}. 
We use an internal release of the \textsc{SGX-Spark} framework.

\subsection{Experiment configurations}

We compare the results of 3 different systems (or execution modes): the vanilla Spark (our baseline), the \sgxspark system with enclaves disabled (\textit{i.e.} collaborative JVMs communicating over SHM which run outside the SGX enclaves) and \sgxspark with enclaves enabled. The latter mode is the one the proposed system runs in. The current implementation of \sgxspark (still under development) does not provide support for Spark's \texttt{Streaming Context} inside enclaves. 
To overcome this temporary limitation, we evaluate the \texttt{SDNN} and \texttt{Identity} algorithms in batch and stream mode. 
For the former, all three different execution modes are supported.
For the latter, we present estimated results for \sgxspark with enclaves enabled, basing the computation time on the batch execution times and the additional overhead against the other modes.
The algorithms are fed with a data file or a data stream, respectively.
In the streaming scenario, an output file is generated every ten seconds.
In a multi-client scenario, each client has a separated data stream (or file) and consequently a different result file.
A streaming execution consists of 5 minutes of the service operating with a specific configuration.
We execute our experiments 5 times and report average and standard deviations.

\textbf{Metrics.} 
To assess performance, scalability, and efficiency, we consider average batch processing times for streaming jobs, and elapsed times for batch executions. Note that we mention \textit{batch} in two different contexts: batch execution (one static input and static output) and streaming batches. Spark Streaming divides live input data in chunks called \emph{batches} determined by a time duration (10 seconds in our experiments). The time it takes the engine to process the data therein contained is denoted as batch processing time. In order to obtain all batch processing times, we rely on the internal Spark's REST API~\cite{spark-rest-api}. Since the GET request fetches the historic of batch processing times for the running job, one single query right before finishing the execution provides all the sufficient informations for our computations. In order to obtain the elapsed times for batch executions, a simple logging mechanism suffices.

\textbf{Workload.}
The clients inject streams as cardiac signals, as shown earlier (\S\ref{subsec:background:med}). Each signal injects a modest workload into our system (230 - 690 bytes per minute). Hence, to assess the efficiency and the processing time as well as to uncover possible bottlenecks, we scale up the output rate of these signals with the goal of inducing more aggressive workloads. We do so in detriment of medical realism, since arbitrary input workloads do not relate to any medical situation or condition. Table~\ref{tab:eval:inputs} shows the variations used to evaluate the various execution modes.

\begin{table}[t]
    \centering
    \caption{Different input loads used for Batch Executions (BE) and Streaming Executions (SE). We present the sample rate they simulate (\textit{i.e.} how many RR intervals are streamed per second) and the overall file or stream size (Input Load). \label{tab:eval:inputs}}
    \begin{tabular}{L{2.5cm}C{5.5cm}C{3.85cm}}
        \toprule
        \textbf{Experiment} & \footnotesize{\textbf{\texttt{s\_rate} (samples / sec)}} & \textbf{Input Load}\vspace{-2pt}\\[0pt] \midrule 
        BE - Small Load & $\lbrace 44, 89, 178, 356, 712, 1424 \rbrace $ & $\lbrace 1, 2, 4, 8, 16, 32 \rbrace$ kB \\[1pt] 
        SE - Small Load & $\lbrace 44, 89, 178, 356, 712, 1424 \rbrace$ & $\lbrace 1, 2, 4, 8, 16, 32 \rbrace$ kB / sec\\[1pt] 
        BE - Big Load & $\lbrace 44, 89, 178, 356, 712, 1424 \rbrace * 1024$ & $\lbrace 1, 2, 4, 8, 16, 32 \rbrace$ MB \\[1pt] 
        SE - Big Load & $\lbrace 44, 89, 178, 356, 712, 1424 \rbrace * 1024$ & $\lbrace 1, 2, 4, 8, 16, 32 \rbrace$ MB / sec\\[1pt] 
        \bottomrule
        %\bottomrule
    \end{tabular}
	%\vspace{-22pt}
\end{table}

\subsection{Results}

\begin{figure}[t]
    \centering
    \includegraphics[width=\textwidth]{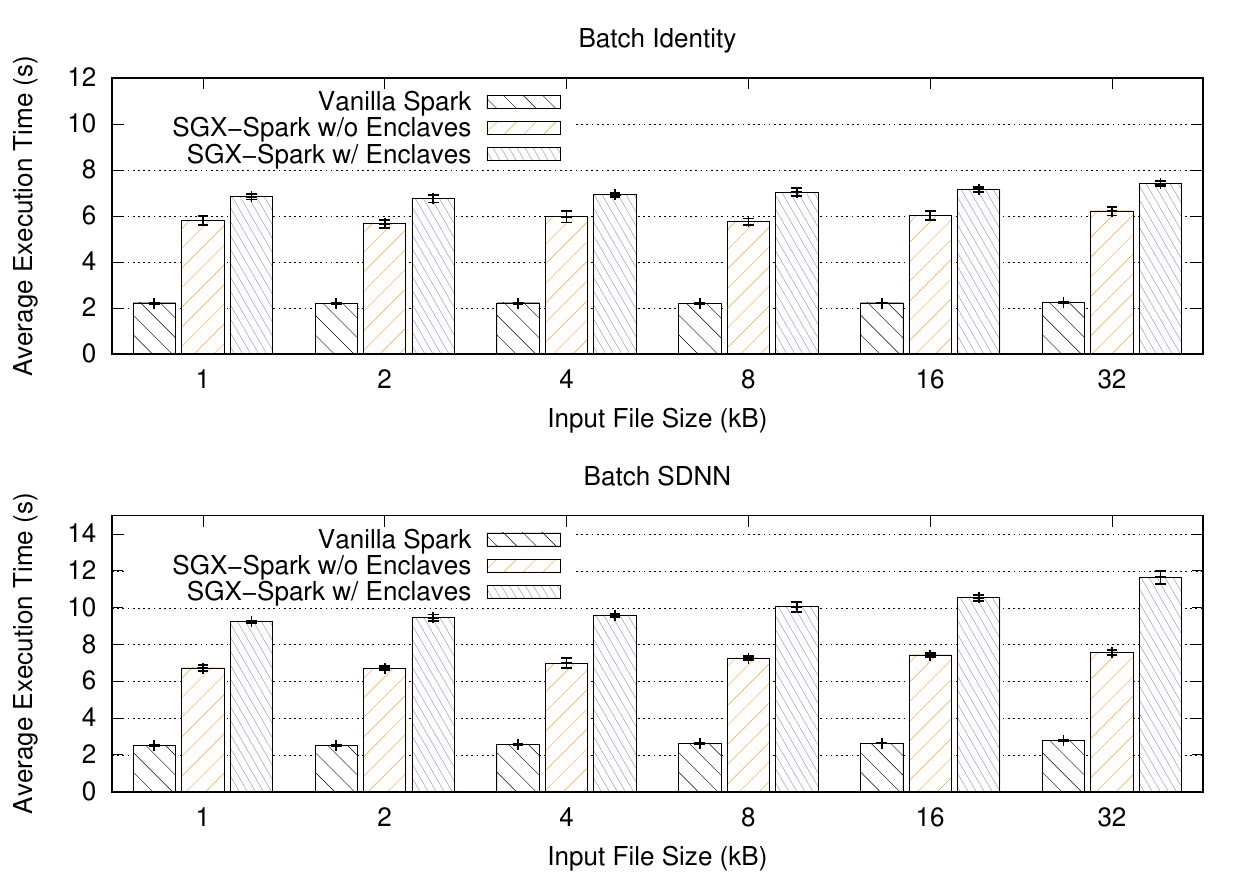}
    \includegraphics[width=\textwidth]{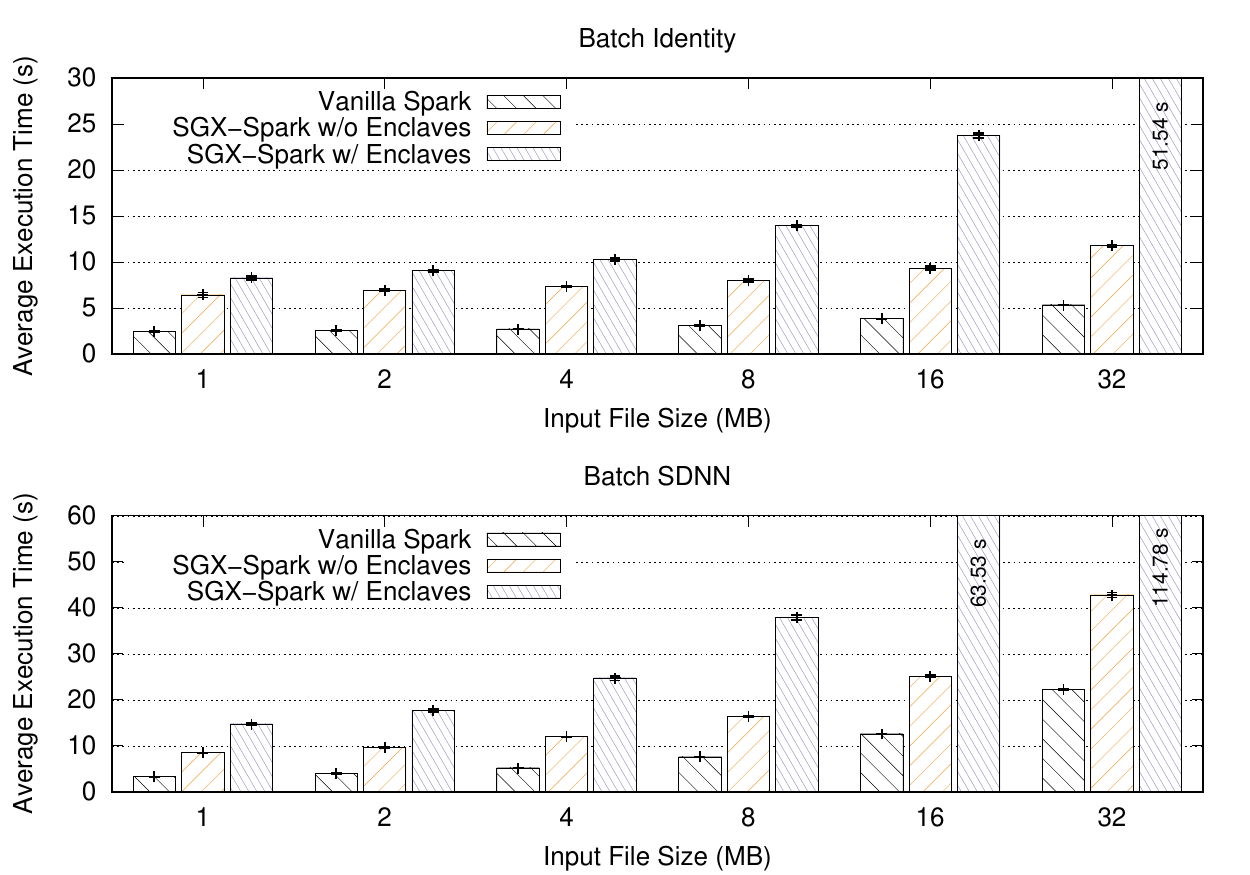}
    \caption{Evolution of the average elapsed time, together with its standard deviation, as we increase the size of the input file. We compare the three different execution modes for each algorithm. Mode \sgxspark w/ enclaves is the mode our system runs in. \label{fig:batch-input-size}}
\end{figure}

\textbf{Batch Execution: input file size.}
The configuration for the following experiments is: one client, one master, one driver, one worker, and a variable input file that progressively increases in size. We measure the processing (or elapsed) time of each execution and present the average and standard deviation of experiments with the same configuration. The results obtained are included in Figure~\ref{fig:batch-input-size}.

From the bar plot we highlight that the variance between execution times among same execution modes as we increase the input file size is relatively low. 
However, it exponentiates as we reach input files of 4-8 MB.
We also observe that the slow-down factor between execution modes remains also quite static until reaching the before mentioned load threshold. 
\sgxspark with enclaves, if input files are smaller than 4 MB, increases execution times x4-5 when compared to vanilla Spark and x1.5-2 when compared to \sgxspark with enclaves disabled.
Note that, since a single client in our real use case streams around 230 to 690 bytes per minute, the current input size limitation already enables several concurrent clients. %(see Figure~\ref{fig:clients})

\begin{figure}[t]
    \centering
    \includegraphics[width=\textwidth]{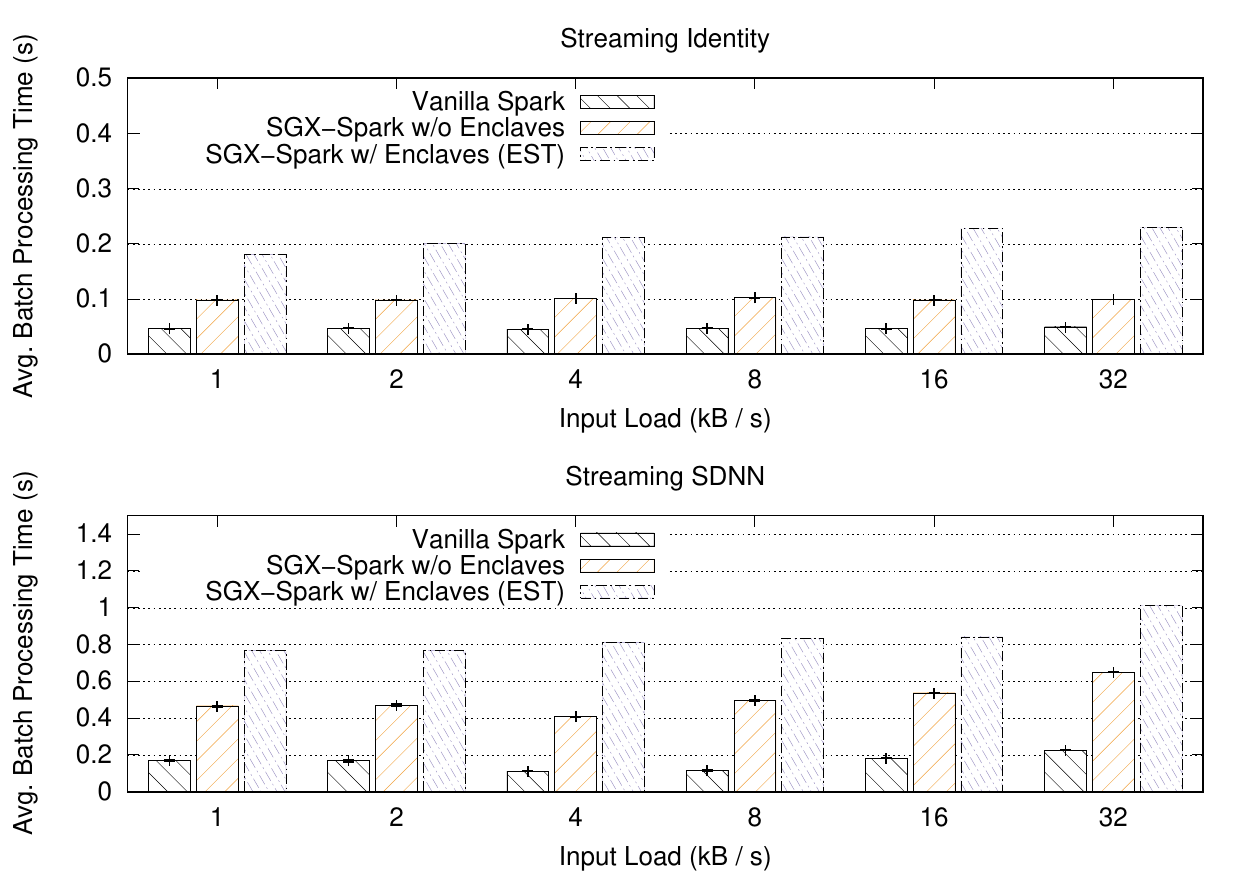}
    \includegraphics[width=\textwidth]{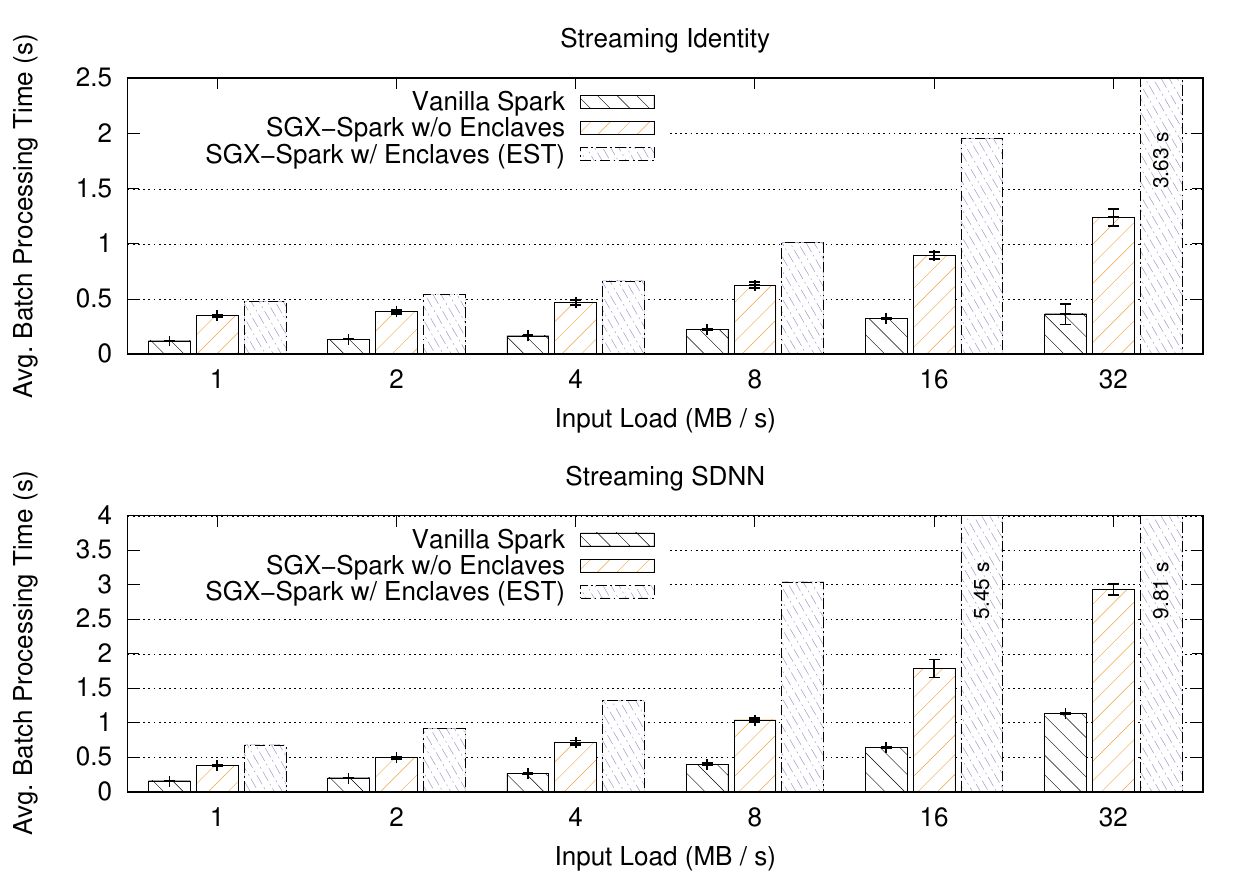}
    \caption{Evolution of the average batch processing time as we increase the the input file size. We compare the results of the three different execution modes. Note that those corresponding to \sgxspark w/ enclaves are estimated basing on the results in Figure~\ref{fig:batch-input-size} and the slow-down with respect to the other execution modes. \label{fig:throughput}}
\end{figure}

\textbf{Streaming Execution: input load.}
As done previously, we scale the load of the data streams that feed the system. We deploy one worker, one driver and one client, query the average batch processing time to Spark's REST API, and present the results for the \texttt{Identity} and \texttt{SDNN} algorithms. Results are summarized in Figure~\ref{fig:throughput}.

We obtain results for vanilla Spark, and \sgxspark without enclaves, and we estimate them for \sgxspark with enclaves.
We observe similar behavior as those in Figure~\ref{fig:batch-input-size}. 
Variability among same execution modes when increasing the input stream size is low until reaching values of around 4 to 8 MB per second.
Similarly, the slow-down factor from vanilla Spark to \sgxspark without enclaves remains steady at around x2-2.5 until reaching the load threshold.
As a consequence, it is reasonable to estimate that the behavior of \sgxspark with enclaves will preserve a similar slow-down factor ($\times$4-$\times$5) when compared with vanilla Spark in streaming jobs.
Similarly, the execution time will increase linearly with the input load after crossing the load threshold of 4 MB.
Note as well how different average batch processing times are in comparison with elapsed times, in spite of relatively behaving similar. 
The average of streaming batch processing times smoothens the initial overhead of starting the Spark engine, and data loading times are hidden under previous batches' execution times.

%% file: sections/related_work.tex
% !TEX root = ../main.tex
\vspace{-15pt}
\section{Related Work}\label{sec:related-work}
\vspace{-10pt}
Stream processing has recently attracted a lot of attention from academia and industry~\cite{Koliousis2016,Miao2017,Venkataraman2017}. Apache Spark~\cite{spark-whitepaper} is arguably the de-facto standard in this domain, by combining batch and stream processing with a unified API~\cite{zaharia2016apache}. Apache Spark SQL~\cite{spark-sql} allows to process structured data by integrating relational processing with Spark's functional programming style. Structured streaming~\cite{structured-streaming} leverages Spark SQL and it compares favorably against the discretized counterpart~\cite{spark-streaming}. However, the former lacks security or privacy guarantees, and hence it was not considered. The proposed system relies on \sgxspark, as it directly extends Spark with SGX support.

Opaque~\cite{opaque} is a privacy-preserving distributed analytics system. It leverages Spark SQL and Intel SGX enclaves to perform computations over encrypted Spark DataFrames. In \texttt{encryption} mode, Opaque offers security guarantees similar to the proposed system. However, \emph{(1)} the Spark master must be co-hosted with the client, a scenario not supported by our multi-client setting and \emph{(2)} it requires changes to the application code. In \texttt{oblivious} mode, \emph{i.e.}, protecting against traffic pattern analysis attacks, it can be up to $46\times$ slower, a factor not tolerable for the real-time analytics in our setting. SecureStreams~\cite{securestreams} is a reactive framework that exploits Intel SGX to define dataflow processing by pipelining several independent components. Applications must be written in the Lua programming language, hindering its applicability to legacy systems or third-party programs. \textsc{DPBSV}~\cite{Puthal2015} is a secure big data stream processing framework that focuses on securing data transmission from the sensors or clients to the \textit{Data Stream Manager (DSM)} or server. Its security model requires a PKI infrastructure and a dynamic prime number generation technique to synchronously update the keys. In spite of using trusted hardware on the DSM end for key generation and management, the server-side processes all the data in clear, making the framework not suitable for our security model. 

Homomorphic encryption~\cite{fhe-definition} does not rely on trusted execution environments and offers the promise of providing privacy-preserving computations over encrypted data.
While several works analyzed the feasibility of homomorphic encryption schemes in cloud environments~\cite{mr-crypt,styx}, the performance of homomorphic operations~\cite{gottel2018security} is far from being pragmatic.

Further, for the specific problem of HRV analysis, while periodic monitoring solutions exist~\cite{Renevey2018}, they are focused on embedded systems. As such, since they off-load computation to third-party cloud services, these solutions simply overlook the privacy concerns that the proposed system considers.

To the best of our knowledge, there are no privacy-preserving real-time streaming systems specifically designed for  medical and cardiac data. The proposed system fills this gap by leveraging Intel SGX enclaves to compute such analytics over public untrusted clouds without changing the existing Java- or Scala-based source code.

%% file: sections/future_work.tex
% !TEX root = ../main.tex
\section{Future Work}\label{sec:futurework}
The current prototype can be improved along several dimensions. First, we envision to support clients running inside ARM TrustZone: this TEE is widely available in low-power devices (\emph{e.g.}, Raspberry PI), hence makes an ideal candidate to reduce the TCB in the client-side of the architecture. Second, we intend to improve the plug-in mechanism for additional analysis of the data, as currently a given algorithm is set at deploy-time, while it is expected to load/unload those at runtime. Thirdly, we intend to study the cost of deployment of such system over public cloud infrastructures such as AWS Confidential Computing.

%% file: sections/conclusion.tex
% !TEX root = ../main.tex
\section{Conclusion} \label{sec:conclusion}
We presented a stream-processing architecture and implementation that leverage Spark-SGX to overcome privacy concerns of deploying such systems over untrusted public clouds. Its design allows to easily scale to different types of data generators (\emph{e.g.}, the clients). The processing components that execute on the cloud rely on \sgxspark, a stream processing framework that can executes Spark jobs within SGX enclaves. Our evaluation shows that for typical signal processing, despite an observed overhead of $4\times-5\times$ induced by the current experimental version of \sgxspark, the performance is still practical. This suggests that it will be possible in a near-future to deploy such systems on a production-ready environment with performances that can easily satisfy even strict Service Level Agreements, while keeping maintaining the costs to use the cloud infrastructure reasonable. We intend to release the code as open-source.

%% file: sections/acks.tex
% !TEX root = ../main.tex

\section{Acknowledgements}\label{sec:acks}
We are grateful to the members of the LSDS Team~\footnote{\url{https://lsds.doc.ic.ac.uk/}} at Imperial College London to have provided us early access to \sgxspark.

\vfill